\documentclass[12pt, preprint]{aastex}
\usepackage{graphicx}

\newcommand{\phunit}{photons cm$^{-2}$ sr$^{-1}$ s$^{-1}$ \AA$^{-1}$}
\newcommand{\ergunit}{ergs cm$^{-2}$ sr$^{-1}$ s$^{-1}$}
\newcommand{\fuse} {{\it FUSE}}
\newcommand{\iras} {{\it IRAS}}
\newcommand{\voyager} {{\it Voyager}}

\begin{document}

\title{Observations of the Diffuse FUV Background with \fuse
\thanks{Based on observations made with the
          NASA-CNES-CSA Far Ultraviolet
          Spectroscopic Explorer. FUSE is operated
          for NASA by the Johns Hopkins University
          under NASA contract NAS5-32985.} }
\author{Jayant Murthy}
\affil{The Indian Institute of Astrophysics}
\affil{Koramangala, Bangalore 560 034}
\email{jmurthy@yahoo.com}
\and
\author{David. J. Sahnow}
\affil{Dept. of Physics and Astronomy}
\affil{The Johns Hopkins University, Baltimore, Md. 21218}
\email{sahnow@pha.jhu.edu}

\begin{abstract}
We have used observations taken under the \fuse\ S405/S505 channel realignment program to explore the
diffuse FUV (1000 - 1200 \AA) radiation field. Of the 71 independent locations in that program, we have 
observed a diffuse signal in 32, ranging in brightness from 1600 \phunit\ to a maximum of
$2.9 \times 10^5$ \phunit\ in Orion. The \fuse\ data confirm that the diffuse FUV sky is
patchy with regions of intense emission, usually near bright stars, but also with dark regions,
even at low Galactic latitudes. We find a weak correlation between the FUV flux and the 100
\micron\ ratio but with wide variations, perhaps due to differences in the local radiation
field.
\end{abstract}


\section{Introduction}

The diffuse background from the ultraviolet (UV) to the infrared (IR) is an important tracer of the 
interstellar dust and most of our knowledge of the large scale distribution of the dust has come from 
missions such as \iras\ and {\it COBE} \citep[see, for example, ][]{S97}.
Scattering in the UV is complementary to the IR emission and the combination of the two
can lead to a unique determination of the interstellar dust parameters. Unfortunately, there have 
been few observations of the diffuse UV radiation field and those have been, to a large
degree, controversial as indicated by the conflicting reviews by \citet{B91} and \citet{H91}. In
the far-ultraviolet (FUV - below 1200 \AA) band, which we address in this paper, the only significant
body of observations comes from \citet{M99}. They
used the ultraviolet 
spectrographs (UVS) aboard the two \voyager\ spacecraft finding that the FUV sky was very patchy with both
dark and bright regions. 

In this work, we have used serendipitous observations from the {\it Far Ultraviolet 
Spectrographic Explorer} (\fuse) under the S405/505 program 
to further probe the diffuse FUV sky. Although \fuse\ cannot match the
sensitivity of the \voyager\ UVS for observations of diffuse sources because of its relatively small 
field of view, we have, nevertheless, found many locations that do indeed have a strong enough signal
to be detected by \fuse. We concentrate here on presenting the 
overall results from our study  and will discuss individual locations in detail in subsequent papers. 

\section{Observations and Data Analysis}

The \fuse\ spacecraft and mission has been described by \citet{M00} and by \citet{S00}. The instrument
consists of four coaligned optical channels, two of which are coated with silicon carbide (SiC)
and two with lithium fluoride (LiF) over aluminum providing coverage over the spectral range from 905 -- 1187 \AA. 
Observations may be made through any of 3 apertures: the LWRS ($30\arcsec\ \times\ 30\arcsec$)
aperture; the MDRS ($4\arcsec\ \times\ 20\arcsec$) aperture; and the 
HIRS ($1.25\arcsec\ \times\ 20\arcsec$) aperture. In principle, extended radiation will be 
visible in all 4 channels and through all the apertures but, in practice, only the brightest sources
can be detected in any other than the LiF LWRS channel. \fuse\ was 
launched on June 24, 1999 into a low Earth orbit (LEO) by a Delta II rocket and has been observing
astronomical targets, mostly point sources, since then.

The S405/505 program is intended to allow the \fuse\ spectrographs to thermalize prior to a channel 
realignment. As such, these pointings are generally observations of blank sky near one of a number of
alignment stars with exposure times on the order of a few thousand seconds. The complete list of
pointings 
is available from the MAST archive at STScI (http://archive.stsci.edu) and, of those,
we have examined all that were available
before Sept 1, 2003. We downloaded the raw data and processed them using the standard CalFUSE
pipeline \citep[v2.4, ][]{D02} with two major modifications. 

The standard \fuse\ observation consists of a number of different exposures including both 
the ``DAY'' and the ``NIGHT'' part of the orbit. Because of the faintness of the diffuse background,
we used only the ``NIGHT'' photons, thereby eliminating most of the airglow
lines other than the Lyman lines of atmospheric hydrogen. There may still be residual amounts
of the \ion{O}{1} lines around 1040 \AA\ and the \ion{N}{1} lines at 1134 \AA\ but these are generally
weak and will not be a significant contributor to the continuum emission reported on here \citep{F01}.
Finally, we
combined the different exposures (using the program {\it ttag\_combine.c} available as part of
the standard \fuse\ distribution). 

\clearpage

\begin{figure}[t]
\figurenum{1}
\plotone{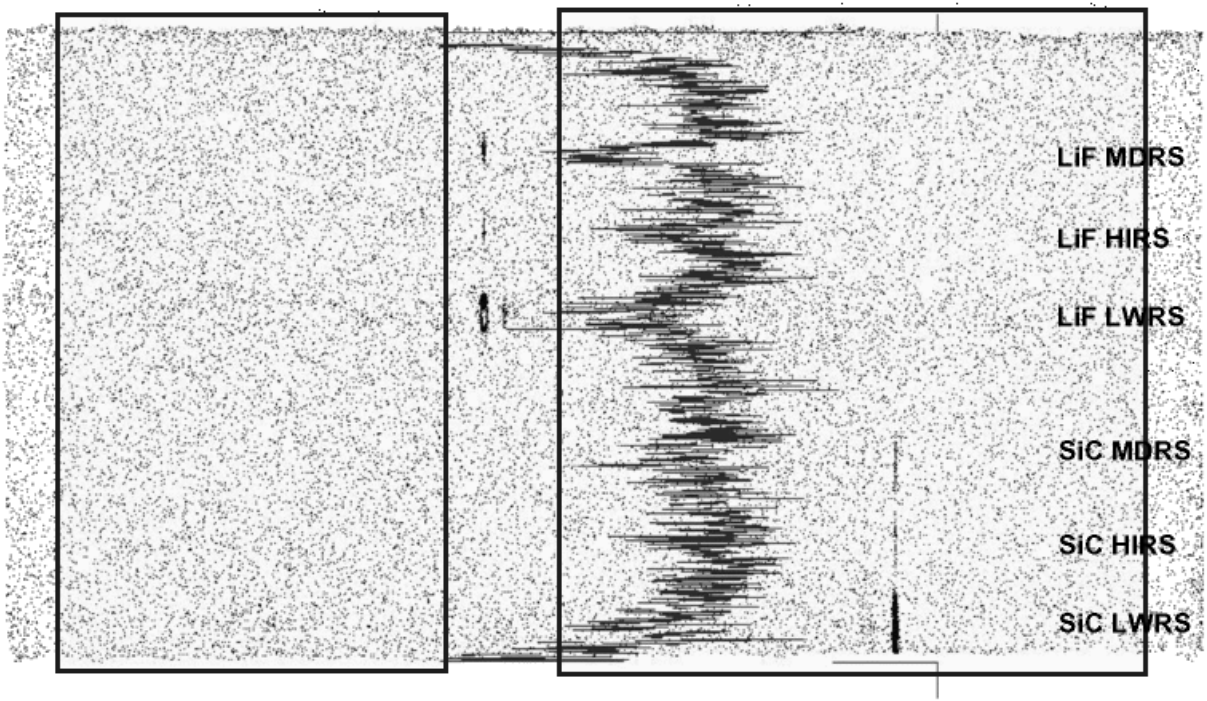}
\caption{We have shown an image of the 1A detector segment from S4050201, a 5626 second
observation of blank sky offset by 90\arcsec\ from a white dwarf. The LiF apertures
are imaged onto the top half of the image and the SiC on the bottom half with the strong terrestrial
Ly$\beta$ line seen as the strongest line in each of the 6 apertures. The image scale can be derived from
the sides of the boxes which are at column numbers 1100, 6000, 7500, and 15000 from left to right.
Superimposed on the image is a cut across the image 
with the data collapsed in the spectral direction over one of the wavelength bands 
(columns 7500 -- 15000; see Table \ref{tbands}) in which the enhancements due
to the diffuse signal in the different apertures can be clearly seen. Note that the defined
bands (shown by the two large boxes) exclude the strong geocoronal emission lines seen in the image. A detailed analysis (see below)
shows that the emission has a flux of 3300 $\pm$ 1400 \phunit\ at a 90\% confidence level and
a mean wavelength of 1058 \AA. The rise in signal at the top and bottom of the active area is due to edge
effects in the microchannel plates.}
\label{ttagimg}
\end{figure}

\clearpage

We have found that the standard background subtraction considerably overestimates the 
instrumental background for the faint extended sources observed in this program and so, instead,
empirically estimated the background
from the counts in the detector just off the aperture and subtracted that from the spectrum. In most
of the targets in this program, the signal was too faint to obtain a useful spectrum even though a diffuse
continuum was clearly apparent to the eye. We, therefore,
used the {\it ttgd} image of the detector plane \citep[Step 19 from the CALFUSE Pipeline 
Reference Guide 2002:][]{D02} and integrated over bands selected to avoid the airglow
lines (Table \ref{tbands}). This is illustrated in Fig. \ref{ttagimg} where we have shown an
image of one of the detector segments (1A) for the S4050201 observation. The two bands (Rows 1 and 2) 
of Table \ref{tbands} are shown as large boxes on either side of the LiF LWRS Ly $\beta$ feature.

The enhancement due to the diffuse continuum is readily visible in the LiF LWRS
aperture, and, upon integration over the bands, all the apertures stand out over the background. This
is clearly shown in Fig. \ref{ttagimg} where we have superimposed a cut across the image in which the
data have been collapsed in the spectral direction over the right hand box of the Figure 
(columns 7500 - 15000 from Row 2 of Table \ref{tbands}). Although, in principle, a diffuse
signal will be visible in all the apertures, we have used only the data from the LiF LWRS aperture
because its throughput is so much greater than the others.
Similarly, the data from the 2B detector do not add any any value to the diffuse sky determination
because of its much lower sensitivity.

\clearpage

\begin{deluxetable}{llll}
\tablecaption{Bands used for background extraction}
\tablenum{1}
\tablewidth{0pt}
\tablehead{
\colhead{Number} &\colhead{Detector} & \colhead{Columns} & \colhead{Wavelengths (\AA)}}
\startdata
1 & LiF 1A & 1100 -- 6000 & 987.08 -- 1020.77 \\
2 & LiF 1A & 7500 -- 15000 & 1034.84 -- 1081.37 \\
3 & LiF 1B & 2000 -- 7000 & 1100.28 -- 1133.69 \\
4 & LiF 1B & 7000 -- 14000 & 1133.69 -- 1180.07 \\
5 & LiF 2A & 2000 -- 7000 & 1175.32 -- 1141.97 \\
6 & LiF 2A & 9000 -- 14000 & 1128.57 -- 1095.03 \\
\enddata
\label{tbands}
\end{deluxetable}

\clearpage

\begin{figure}[t]
\figurenum{2}
\plotone{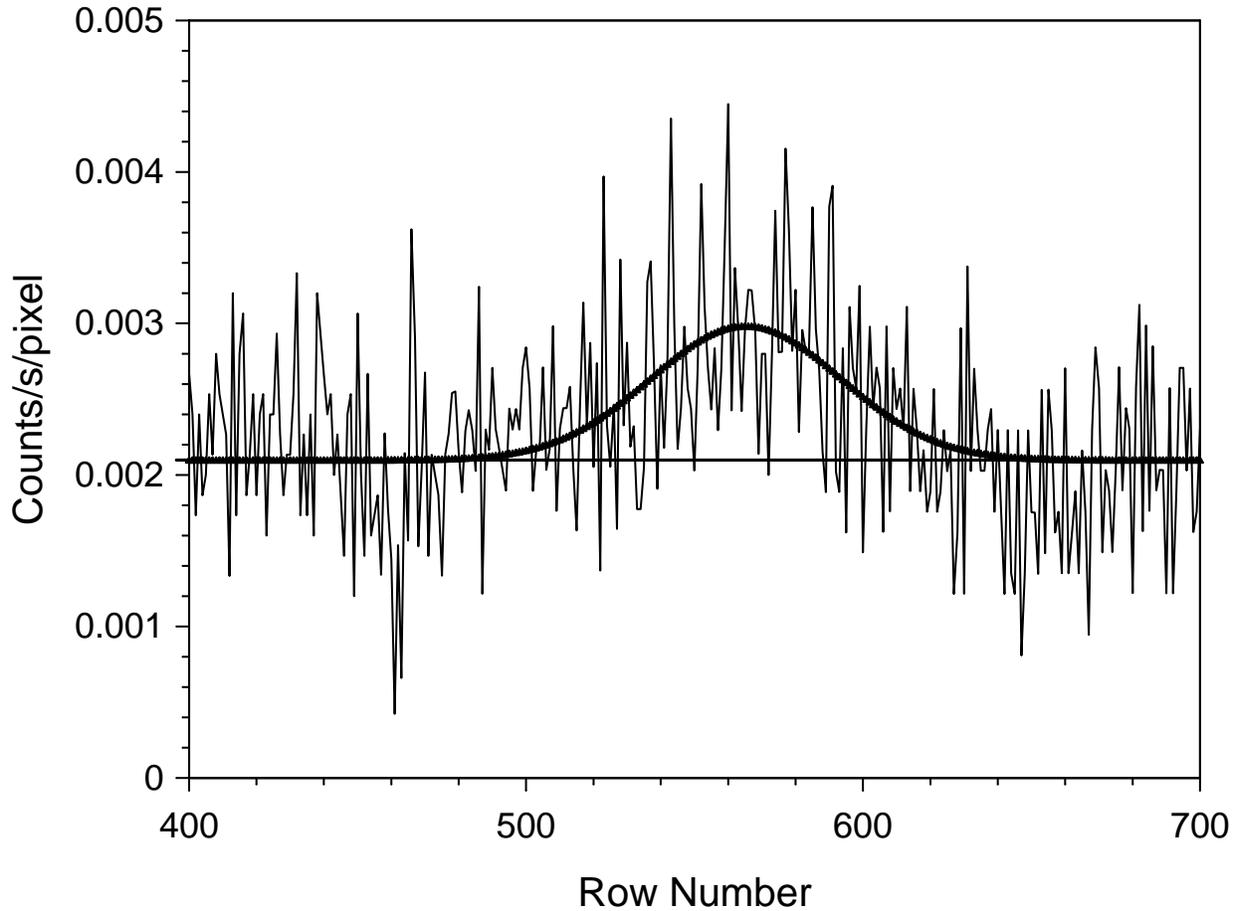}
\caption{Plotted here is a cut through the detector plane from S4050201 (Fig. \ref{ttagimg}). We have
integrated between columns 7500 and 15000 (Row 2 of Table \ref{tbands}) which avoids the Ly $\beta$ airglow lines.
The LiF LWRS aperture stands out clearly above the background and
we have fit the signal with a Gaussian, 
shown as a dark line. As mentioned above the level of emission here is 3300 $\pm$ 1400 \phunit.
The noise level in the data sets a detection limit on the order of 2000 \phunit.}
\label{profile}
\end{figure}

\begin{figure}[t]
\figurenum{3}
\plotone{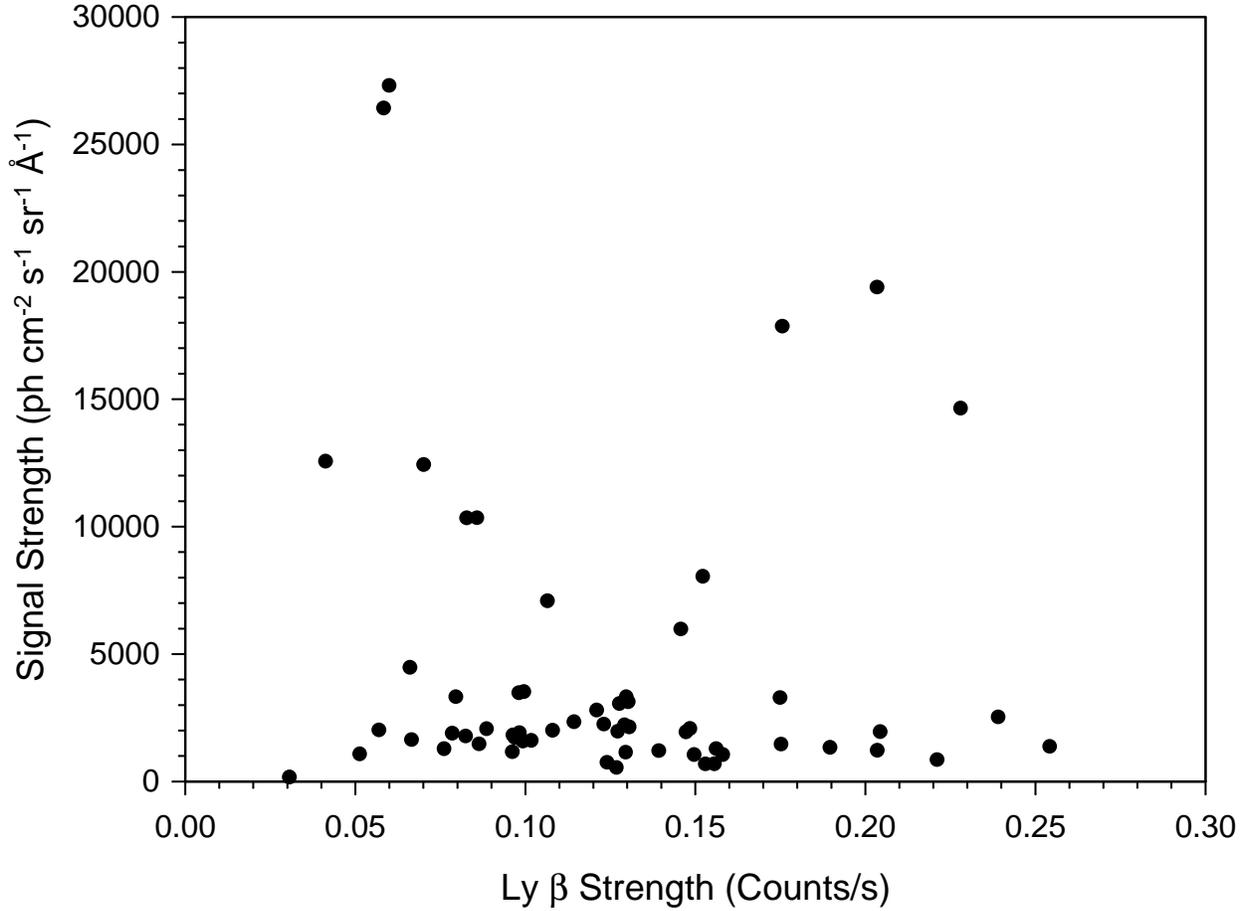}
\caption{We have plotted the derived diffuse continuum signal versus the total
number of counts under the Ly $\beta$ line. If scattering from geocoronal lyman lines is a 
significant contaminant of the observed spectrum, we would expect a correlation.}
\label{lyb_fig}
\end{figure}

\clearpage

As we have seen from Fig. \ref{ttagimg}, the emission in the LiF LWRS aperture, in particular, stands
out from the background and we have replotted the signal in the immediate neighbourhood of the aperture in
Fig. \ref{profile}. We have then fit this profile with a Gaussian (plus a background)
with uncertainties defined 
by the root mean square deviations adjacent to the aperture and found 90\% confidence limits on the
level of the diffuse background using the procedure of \citet{L76}. Those targets in which
we have observed a diffuse astronomical signal are listed in Table \ref{pos_det}.
The sensitivity limit of the \fuse\ spectrographs to diffuse radiation is on the order of
about 2000 \phunit\ and so the null detections are not interesting.

This procedure is tantamount to assuming that the instrumental background is the
same in the aperture as off. There are several instrumental effects which may affect this, of which
the most likely to be a problem is scattering in the spectral direction from the Lyman lines of
atmospheric \ion{H}{1}. We have tested for this by plotting the observed signal against the counts
under the Ly $\beta$ line (Fig. \ref{lyb_fig}) for a representative sample and found
no correlation between the astronomical and geocoronal lines. While there are other possibilities,
we have found no evidence for any 
aperture dependent effects in our null detections --- the signal is flat over the entire detector. 
Perhaps the strongest argument for the quality of our background subtraction comes from the 
excellent agreement between different segments and different observations, separated in time by as much
as a year and a half, in all of which the derived
background agrees within the error bars. The only exception is S4055401 in which the background
derived from segment 2A is much higher than the others. An examination of the raw data shows that
the count rate is much higher at the beginning of each exposure suggesting contamination from daylight photons.

\clearpage

\begin{figure}[t]
\figurenum{4}
\plotone{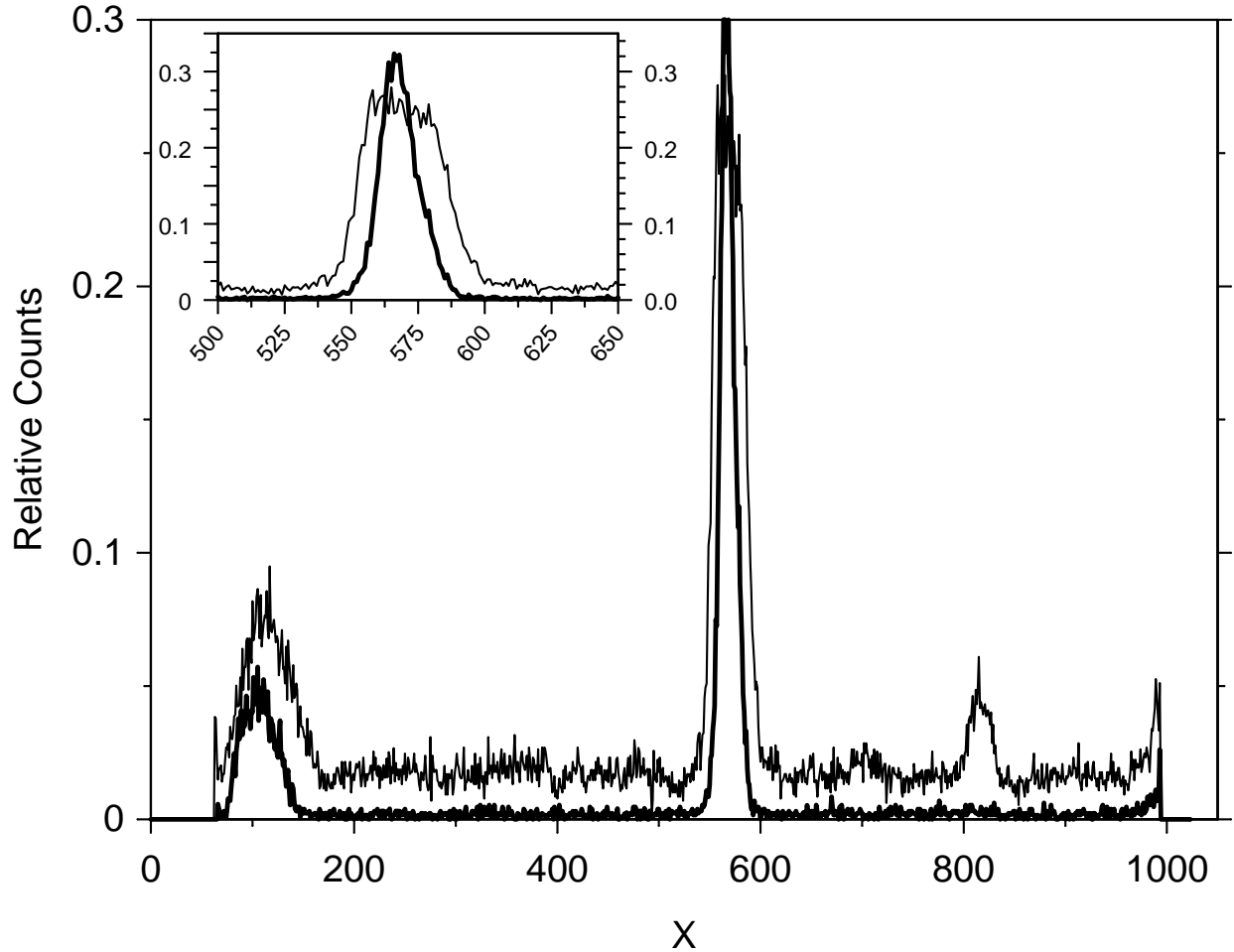}
\caption{A vertical cut through the image shows that the PSF for a star (lower line) 
is much narrower than that
for a diffuse source, in this case an observation in Orion. Note that the emission from the
star is only visible in the LWRS while the diffuse radiation is apparent in the LiF MDRS also. An
expanded view of the profile in the vicinity of the LiF LWRS aperture is shown in the inset.
}
\label{ps_comp}
\end{figure}

\clearpage

The signal levels are so low that stellar contamination might be a serious problem. The sensitivity 
limit of 2000 \phunit\ corresponds to an unreddened B star of about $16^{th}$ magnitude in V.
We have examined each of the fields using the Digital Sky Survey (DSS) plates  
and have rejected those few fields in which there were stars that were bright enough to possibly 
affect our determination of the diffuse background. Most of these were in the SMC or LMC where there
are many hot bright stars, some of which did fall in the \fuse\ FOV. Another test of stellar 
contamination comes from the much broader spread for a diffuse source, which fills the aperture,
as opposed to a point source
(Fig. \ref{ps_comp}) and we have confirmed that the spread for those sources identified as diffuse 
is really larger than that of a star. In practice, there are few unreddened early-type stars in the 
sky and any stellar contribution in the
\fuse\ bandpass will be heavily depressed because of interstellar extinction. Finally, we
excluded those observations in which the pointing was particularly poor.

We have also considered whether scattering from the nearby alignment star, which may be quite 
bright in the FUV, can contribute to the diffuse signal. The \fuse\ instrument team
has studied the scattered light from $\gamma$ Cas and found that the scattered light at a 
distance of 90\arcsec\ from the star is on the order of 7 $\times\ 10^{-6}$ the stellar
flux (personal communication: B.-G. Anderson 2003). Even the brightest stars in our sample, with
an observed intensity of $10^{-11}$ 
\ergunit, will not contribute more than 200 \phunit\ to the signal, much less than our sensitivity
limit.

\section{Results}

\clearpage

\begin{figure}
\plotone{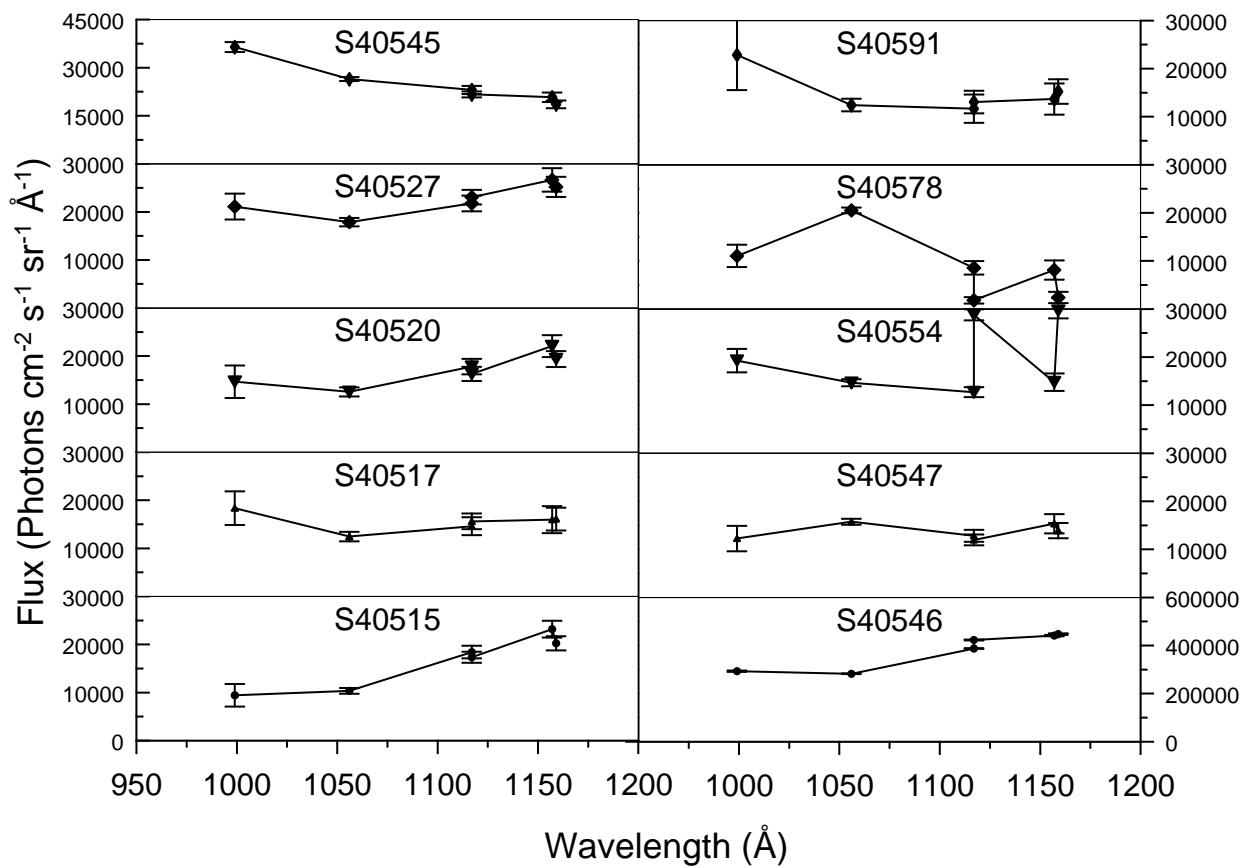}
\figurenum{5}
\caption{We have plotted the flux extracted from each of the bands listed in Table \ref{tbands}
for each of the positive detections in our data base. These include data from all the detectors
except for 2B (for which the effective area was much less than for the other detectors). Note the 
generally excellent agreement in fluxes between different segments.
The two brightest spectra (S40545 and
S40546) are both of targets in Orion. Differences in the spectra may reflect different local
radiation fields.}
\label{fuse_spectra}
\end{figure}

\begin{figure}
\plotone{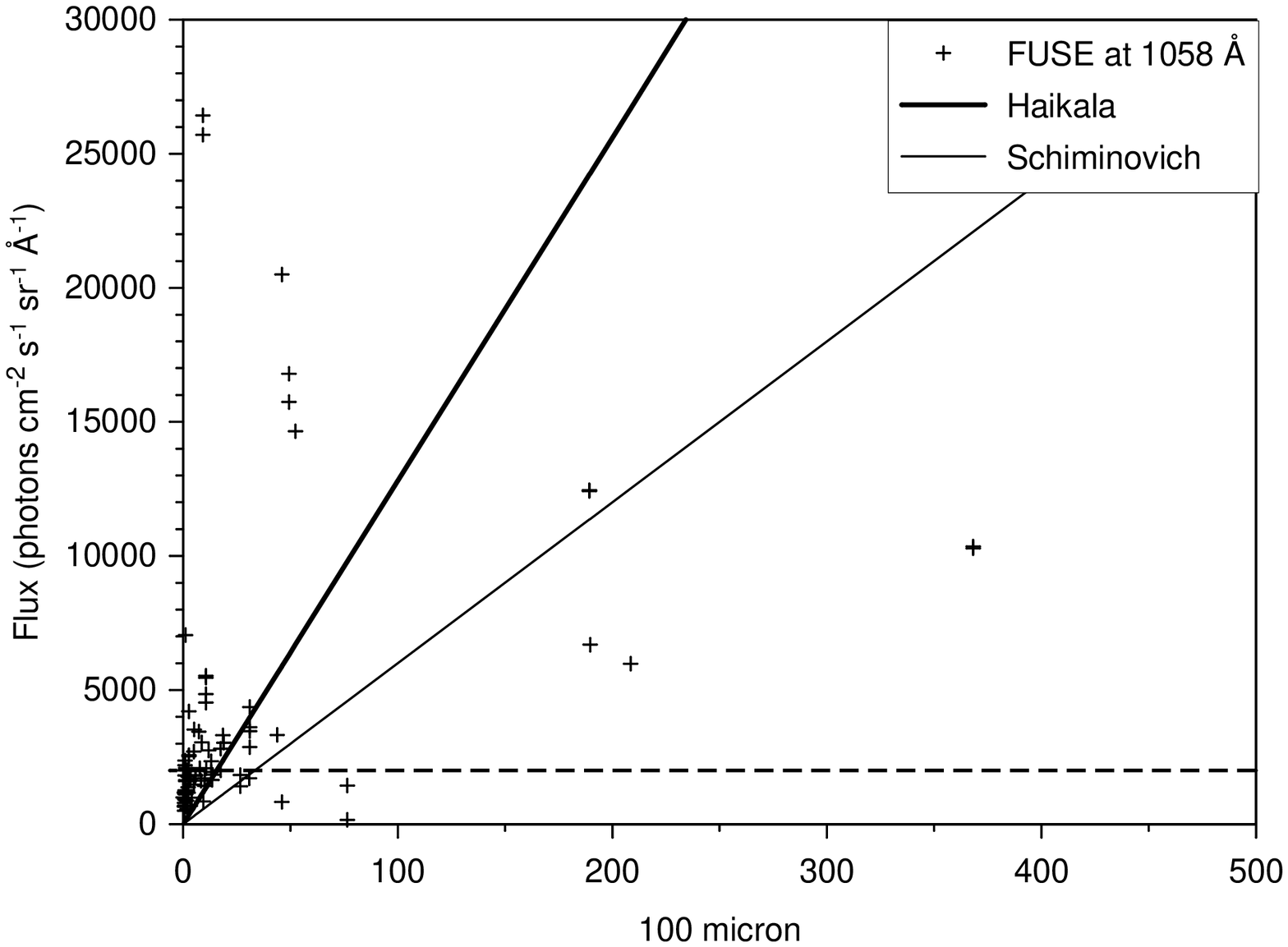}
\figurenum{6}
\caption{
The FUV/IR ratio shows a general trend of increasing FUV flux (at 1058 \AA; Table \ref{tbands} second
row) with increasing
100 \micron\ flux but with wide variations in the actual ratio. The two lines indicate the
NUV/IR ratios obtained by \citet{H95} for an isolated high latitude cirrus cloud (dark line)
and by \citet{S01} for the high latitude diffuse background. We have not
plotted the brightest of our targets --- S40546 in Orion --- which has an observed surface brightness
of $2.9 \times\ 10^5$ \phunit\ in the FUV and 2000 MJy sr$^{-1}$ at 100 \micron.
}
\label{fuse_ir}
\end{figure}

\clearpage

Of the 107 total observations (71 independent targets) in the S405/505 program (to our cutoff date), we have identified 
45 (32 independent locations) as unquestionable detections of a diffuse astronomical signal.
These positive detections are listed in Table \ref{pos_det} and range in 
strength from 1600 \phunit\ to a maximum of $3 \times\ 10^5$ \phunit\ (in the Orion nebular region). The 
brightest of these are plotted in Fig. \ref{fuse_spectra} and show a variety of spectral shapes,
perhaps indicative of the local radiation field. For instance, the scattered spectrum for S40546 ---
a field in Orion --- is very similar to that of the nearby star HD36981 \citep{M04}.
We will discuss each of the individual regions in subsequent papers and concentrate
on the global distribution of the diffuse background in this work. Images and further description
of each of the fields may be found at 
http://www.iiap.res.in/personnel/murthy/projects/fuse/FUSE\_background\_analysis.html.

There have been three studies of the UV/100 \micron\ correlation in the near UV. \citet{H95}
found a ratio of 128 \phunit\ (MJy sr$^{-1}$)$^{-1}$ for an isolated cirrus cloud at high
galactic latitude using FAUST data; \citet{S01} found a latitude dependent
ratio of between 60 (b $>$ 30\fdg) and 100 (b $>$ 15\fdg) \phunit\ (MJy sr$^{-1}$)$^{-1}$ using
the NUVIEWS instrument; and
\citet{M01} found ratios varying between 30 and 300 \phunit\ (MJy sr$^{-1}$)$^{-1}$ 
in {\it Midcourse Space Experiment} ({\it MSX}) observations around M42 in Orion. Our corresponding data are plotted in Fig. \ref{fuse_ir}
with the flux from the 1B spectrum at an effective wavelength of 1058 \AA\ (mean wavelength
for Band 2 in Table \ref{tbands}) plotted against the 
100 \micron\ flux from \citet{S98}. Although there is a trend of increasing
FUV emission with increasing IR, there is considerable variation in the ratio ranging from
only 28 \phunit\ (MJy sr$^{-1}$)$^{-1}$ near the Wolf-Rayet star HD 92809 (S40515) to 
2800 \phunit\ (MJy sr$^{-1}$)$^{-1}$  near the star HD36487 in Orion (S40545). In fact, this 
variation should not be surprising. The UV signal arises from scattering of the interstellar
radiation field (ISRF) by interstellar dust and so depends heavily on the relative orientation of the
stars and the dust, particularly in the FUV where there are only a relatively small number of
bright stars which dominate the ISRF. On the other hand, the IR emission is due to the thermal emission
from the heated interstellar dust and is not dependent on the direction of the incoming radiation.
Moreover, the optical depth in the UV is much higher than in the IR and saturation effects may be
expected to become important even with low column densities of dust.

\begin{figure}
\plotone{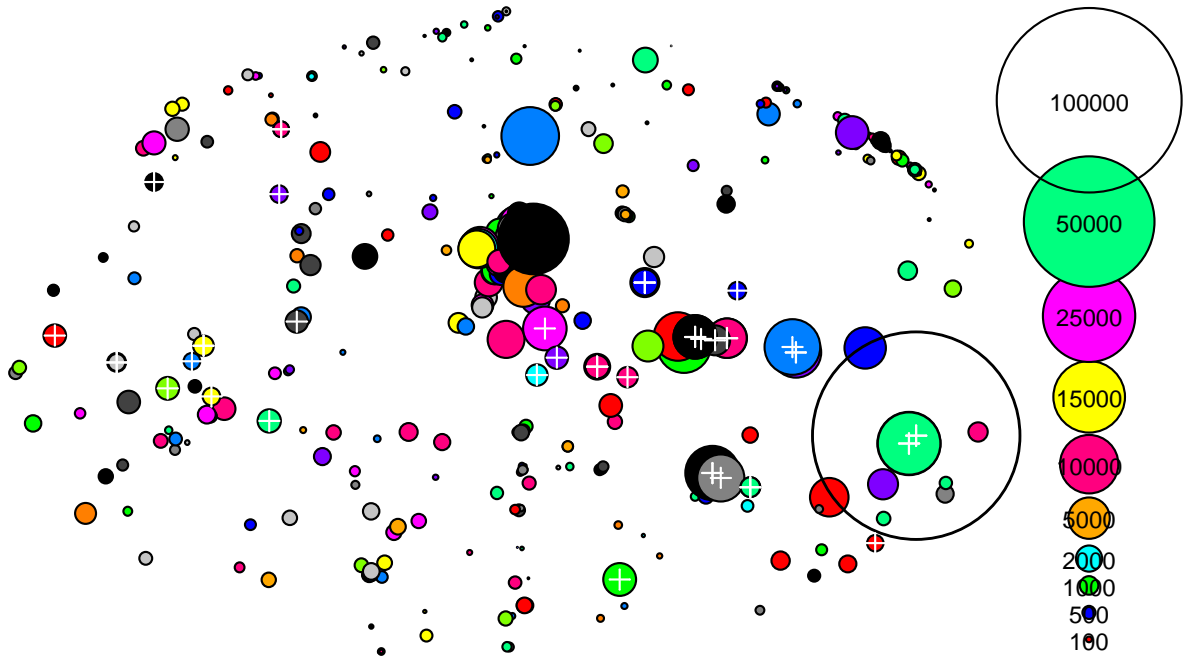}
\figurenum{7}
\caption{
We have combined the 426 Voyager observations of \citet{M99} with the \fuse\ observations
of this paper into an Aitoff projection of the diffuse FUV background with the Galactic centre
in the center of the image and $\pm180\deg$ at the left and right edges, respectively. The area of each
circle is proportional to the observed surface brightness with the large open circle
at the bottom right (Orion) having a brightness of $2.9 \times\ 10^5$ \phunit. (Note that the
colors in the electronic edition are only for clarity and do not reflect the flux.) The circles with
plus signs at the center are the \fuse\ observations presented in this paper while the others are from
\citet{M99}. The brightest regions are Orion in the bottom right and Ophiuchus near the center.
Note that we have not included the null detections of \fuse\ (those of less than about 2000 \phunit\
in strength) which do not place useful
limits on the diffuse signal.
}
\label{fuse_ait}
\end{figure}

As mentioned earlier, the only other major body of observations in the FUV are from observations
made with the Voyager UVX \citep{M99} and we have plotted those data as well as
the data in this work in an Aitoff projection of the sky in Fig. \ref{fuse_ait}. In the Figure, the 
area of the 
circles is proportional to the observed surface brightness with the large open circle in Orion
corresponding to a surface brightness of $2.9 \times 10^{5}$ \phunit. Note that we have not shown
the \fuse\ null detections because the detection limit is too high to be useful. On the other hand, the Voyager null detections
are at a level of only about 100 \phunit\ and are shown in the Figure. Prominent hot spots in 
the map include Orion (near 
the right edge of Fig. \ref{fuse_ait} - Murthy et al. 2004), 
Ophiuchus (near l = 0$\degr$; b = 28$\degr$),
and the Coalsack (l = 305$\degr$; b = 0$\degr$ - Shalima \& Murthy 2004) but it should also be
noted that there are a number of dark regions even at low Galactic latitudes.

It is clear from the data presented in this paper that the intensity and the spectrum of the 
diffuse radiation in the FUV does vary considerably over the sky. Although other studies
\citep[see, for example, ][ and references therein]{S01} have claimed simple correlations between
the diffuse UV radiation and tracers of interstellar dust such as 21 cm \ion{H}{1} column
densities or 100 \micron\ intensities in the NUV, we cannot support such from our data in the FUV.
The optical depth of the interstellar dust is much higher in the FUV and it is possible
that local effects are more important than in the NUV. Thus we will defer modeling of our
results to extract such important quantities as the optical properties of the interstellar dust
grains.

\clearpage

\clearpage

\section{Conclusion}

We have used serendipitous observations of blank sky with the \fuse\ spacecraft to investigate
the diffuse sky background in many areas over the sky. Of the total 71 independent pointings, we have
observed a signal which we can unambiguously attribute to a diffuse background in 32 targets. Considering 
that the S405/505 targets were chosen simply on the basis of a nearby alignment star and
considering that \fuse\ is only sensitive to signals of greater than 2000 \phunit, this is a
surprisingly large percentage. By contrast, in the \voyager\ sample of \citet{M99} only
63 of the total 426 targets have a flux of greater than 2000 \phunit. Of course, in neither case
was an unbiased survey of the diffuse radiation field intended and it is likely that selection effects
play an important role in these ratios.

We have found that there is a trend of increasing FUV flux with the 100 \micron\ flux indicating
that the observed radiation is due to light scattered from the interstellar dust. However, the
ratio between the FUV and the IR varies much more than was found by either \citet{H95} or
\citet{S01} in the NUV. Our targets are in quite different locations in the sky
and it is apparent that local effects --- such as exposure to the intense radiation field in Orion
--- play an important role  in determining the scattering of the stellar radiation. \citet{H95}
derived their ratio for a single isolated cirrus cloud while Schiminovich et al. surveyed a large
fraction of the sky.

If we combine our data with the \voyager\ data of \citet{M99}, we see that the FUV sky is quite
patchy with intensities ranging from upper limits of less than 100 \phunit\ to intense regions as
high as $3 \times\ 10^{5}$ \phunit. These regions are scattered throughout the sky with both
bright and faint regions being found at all latitudes, again suggesting that local effects 
dominate the FUV diffuse radiation field.

\acknowledgements{
We thank the \fuse\ team for much helpful information and discussion. This research has
made use of NASA's Astrophysics Data System and the SIMBAD database operated at CDS, Strasbourg, 
France. The data preented in this paper were obtained from the Multimission Archive at the Space
Telescope Science Institute (MAST). STScI is operated by the Association of Universities for Research
in Astronomy, Inc. under NASA contract NAS5-26555. Support for MAST for non-HST data is provided
by the NASA Office of Space Science via grant NAG5-7584 and by other grants and contracts.
}

\clearpage

\thispagestyle{empty}

\begin{deluxetable}{ccccccccccccc}
\rotate
\tablenum{2}
\tablecaption{Positive detections of diffuse radiation.}
\label{pos_det}
\tabletypesize{\tiny}
\tablehead{
\colhead{Data Set} & \colhead{Target} & \colhead{Ra} & \colhead{Dec} & \colhead{L} &
\colhead{B} & \colhead{Time (s)} & \colhead{1} & \colhead{2} & \colhead{3} & 
\colhead{4} & \colhead{5} & \colhead{6}}
\startdata
S40502/01 & WD0439+466 & 70.8 & 46.7 & 158.5 & 0.5 & 5626 & 7735 $\pm$ 4498 & 3316 $\pm$ 1377 & 3194 $\pm$ 1594 & 3455 $\pm$ 1708 & 2887 $\pm$ 1311 & 3200 $\pm$ 2182 \\
S40506/01 & HD093840 & 162.3 & -46.8 & 282.1 & 11.1 & 12998 & 4850 $\pm$ 3846 & 2088 $\pm$ 751 & 2039 $\pm$ 1063 & 3429 $\pm$ 1531 & 1196 $\pm$ 958 & 2860 $\pm$ 2444 \\
S40507/01 & HD96548 & 166.6 & -65.7 & 292.3 & -60.8 & 7843 & 13639 $\pm$ 5393 & 7051 $\pm$ 1011 & 9408 $\pm$ 2007 & 9441 $\pm$ 2200 & 8141 $\pm$ 1694 & 8979 $\pm$ 1515 \\
S40514/01 & HD163522 & 269.7 & -42.5 & 349.6 & -9.1 & 11001 & 2891 $\pm$ 1606 & 3034 $\pm$ 838 & 3356 $\pm$ 896 & 4910 $\pm$ 1211 & 3269 $\pm$ 1610 & 2977 $\pm$ 1088 \\
S40515/01 & HD92809 & 160.4 & -58.8 & 286.8 & -0.0 & 11303 & 9425 $\pm$ 2358 & 10351 $\pm$ 616 & 18414 $\pm$ 1315 & 23184 $\pm$ 1752 & 20241 $\pm$ 1468 & 17343 $\pm$ 1152 \\
S40515/02 & HD92809 & 160.4 & -58.8 & 286.8 & -0.0 & 3542 & 11878 $\pm$ 5296 & 10283 $\pm$ 1276 & 18356 $\pm$ 2534 & 22032 $\pm$ 2588 & 19354 $\pm$ 2321 & 20185 $\pm$ 2779 \\
S40517/01 & HD104994 & 181.3 & -62.1 & 297.6 & 0.3 & 4154 & 18358 $\pm$ 3483 & 12452 $\pm$ 1006 & 14601 $\pm$ 1866 & 15959 $\pm$ 2801 & 16094 $\pm$ 2380 & 15623 $\pm$ 1615 \\
S40520/01 & HD153426 & 255.3 & -38.2 & 347.1 & 2.4 & 7521 & 14672 $\pm$ 3371 & 12548 $\pm$ 959 & 17809 $\pm$ 1621 & 22077 $\pm$ 2269 & 19387 $\pm$ 1643 & 16278 $\pm$ 1466 \\
S40521/01 & BD+28D4211 & 327.8 & 28.9 & 81.9 & -19.3 & 10730 & 6525 $\pm$ 2248 & 3529 $\pm$ 747 & 2695 $\pm$ 887 & 3675 $\pm$ 1558 & 2970 $\pm$ 2503 & 1549 $\pm$ 821 \\
S40522/01 & HD216438 & 343.0 & 53.7 & 105.7 & -5.1 & 3971 & 1973 $\pm$ 1973 & 1846 $\pm$ 730 & 1174 $\pm$ 838 & 5287 $\pm$ 1991 & 4156 $\pm$ 3067 & 3309 $\pm$ 2832 \\
S40526/01 & HD156385 & 259.9 & -45.6 & 343.2 & -4.8 & 7029 & 3925 $\pm$ 3925 & 3328 $\pm$ 1148 & 6065 $\pm$ 1399 & 8176 $\pm$ 2220 & 6189 $\pm$ 1965 & 3860 $\pm$ 1557 \\
S40527/01 & SK71D45 & 82.8 & -71.1 & 281.9 & -32.0 & 6754 & 21096 $\pm$ 2700 & 17857 $\pm$ 858 & 21757 $\pm$ 1638 & 26638 $\pm$ 2449 & 25196 $\pm$ 2106 & 23063 $\pm$ 1509 \\
S40527/02 & SK71D45 & 82.8 & -71.1 & 281.9 & -32.0 & 3893 & 23156 $\pm$ 3259 & 19405 $\pm$ 1085 & 21430 $\pm$ 2226 & 26891 $\pm$ 2548 & 25993 $\pm$ 2332 & 23756 $\pm$ 1747 \\
S40528/01 & HD187459 & 297.2 & 33.4 & 68.8 & 3.9 & 5129 & 4100 $\pm$ 4100 & 3298 $\pm$ 1810 & 4949 $\pm$ 1376 & 6359 $\pm$ 2227 & 5502 $\pm$ 2025 & 4567 $\pm$ 1510 \\
S40529/01 & HD013268 & 32.9 & 56.2 & 134.0 & -5.0 & 6730 & 3840 $\pm$ 3840 & 2347 $\pm$ 1081 & 2241 $\pm$ 1134 & 2540 $\pm$ 2017 & 2340 $\pm$ 1102 & 2825 $\pm$ 2282 \\
S40529/02 & HD013268 & 32.9 & 56.2 & 134.0 & -5.0 & 5113 & 3955 $\pm$ 3955 & 1949 $\pm$ 1254 & 2980 $\pm$ 1264 & 4374 $\pm$ 2033 & 5107 $\pm$ 3690 & 3300 $\pm$ 3008 \\
S40531/01 & GD50 &  &  & 189.0 & -40.1 & 5915 & 3675 $\pm$ 3675 & 1800 $\pm$ 1026 & 1246 $\pm$ 1142 & 2294 $\pm$ 1719 & 4199 $\pm$ 4199 & 3648 $\pm$ 3648 \\
S40532/01 & BD+532820 & 333.5 & 54.4 & 101.2 & -1.7 & 12077 & 1391 $\pm$ 911 & 2007 $\pm$ 1027 & 2051 $\pm$ 1025 & 4644 $\pm$ 1551 & 2624 $\pm$ 1612 & 3413 $\pm$ 2148 \\
S40532/02 & BD+532820 & 333.5 & 54.4 & 101.2 & -1.7 & 6243 & 2872 $\pm$ 2872 & 2812 $\pm$ 1288 & 2370 $\pm$ 1105 & 3739 $\pm$ 2320 & 2063 $\pm$ 1241 & 3255 $\pm$ 2977 \\
S40544/03 & WD0005+511 &  &  & 116.1 & -10.9 & 2264 & 5746 $\pm$ 5746 & 3451 $\pm$ 1903 & 962 $\pm$ 962 & 2484 $\pm$ 2184 & 6321 $\pm$ 6321 & 2107 $\pm$ 2107 \\
S40545/01 & HD36487 & 82.9 & -7.1 & 210.2 & -21.0 & 18159 & 36434 $\pm$ 1537 & 26436 $\pm$ 608 & 23044 $\pm$ 1264 & 20756 $\pm$ 1465 & 18531 $\pm$ 1191 & 21654 $\pm$ 916 \\
S40545/02 & HD36487 & 82.9 & -7.1 & 210.2 & -21.0 & 8461 & 35871 $\pm$ 2674 & 25708 $\pm$ 830 & 23127 $\pm$ 1651 & 21833 $\pm$ 1607 & 20573 $\pm$ 1891 & 21992 $\pm$ 1429 \\
S40546/01 & HD36981 & 83.8 & -5.2 & 208.8 & -19.3 & 10565 & 293331 $\pm$ 2799 & 282397 $\pm$ 1501 & 387482 $\pm$ 1565 & 441446 $\pm$ 2536 & 447431 $\pm$ 3448 & 423016 $\pm$ 2065 \\
S40546/02 & HD36981 & 83.8 & -5.2 & 208.8 & -19.3 & 5696 & 295623 $\pm$ 4247 & 286129 $\pm$ 1913 & 394862 $\pm$ 1901 & 447044 $\pm$ 2903 & 454589 $\pm$ 3138 & 428521 $\pm$ 2984 \\
S40547/01 & HD72350 & 127.7 & -44.7 & 262.7 & -3.2 & 9375 & 12257 $\pm$ 2641 & 15741 $\pm$ 609 & 12807 $\pm$ 1235 & 15331 $\pm$ 1989 & 13900 $\pm$ 1571 & 11949 $\pm$ 1132 \\
S40547/02 & HD72350 & 127.7 & -44.7 & 262.7 & -3.2 & 4283 & 12018 $\pm$ 4164 & 16791 $\pm$ 974 & 13622 $\pm$ 2076 & 13593 $\pm$ 2430 & 14273 $\pm$ 2084 & 12555 $\pm$ 1752 \\
S40549/02 & NCVZ & 218.2 & 65.0 & 107.0 & 48.8 & 26083 & 3191 $\pm$ 2670 & 1642 $\pm$ 552 & 1166 $\pm$ 605 & 2569 $\pm$ 1219 & 2150 $\pm$ 2150 & 2330 $\pm$ 2330 \\
S40549/03 & NCVZ & 218.2 & 65.0 & 107.0 & 48.8 & 19047 & 2720 $\pm$ 2486 & 1607 $\pm$ 653 & 1194 $\pm$ 848 & 2545 $\pm$ 1109 & 2880 $\pm$ 2880 & 2214 $\pm$ 1505 \\
s40550/01 & Z-Cam & 126.3 & 73.1 & 141.4 & 32.6 & 4497 & 4587 $\pm$ 3619 & 2042 $\pm$ 1238 & 823 $\pm$ 823 & 3391 $\pm$ 2426 & 1717 $\pm$ 1717 & 5073 $\pm$ 5073 \\
s40553/01 & WR42-HD97152 & 167.5 & -61.0 & 290.9 & -0.5 & 11884 & 7747 $\pm$ 2859 & 5983 $\pm$ 792 & 8127 $\pm$ 1421 & 10034 $\pm$ 1726 & 4936 $\pm$ 1401 & 5159 $\pm$ 1432 \\
s40554/01 & Sk-67D111 & 81.7 & -67.5 & 277.8 & -33.0 & 12645 & 19264 $\pm$ 2443 & 14648 $\pm$ 742 & 12697 $\pm$ 1036 & 14773 $\pm$ 1808 & 29656 $\pm$ 1614 & 28805 $\pm$ 1153 \\
S40555/01 & PG1520+525 & 230.5 & 52.4 & 85.4 & 52.4 & 7411 & 3620 $\pm$ 3620 & 2199 $\pm$ 1343 & 1011 $\pm$ 1011 & 2801 $\pm$ 1875 & 1449 $\pm$ 1449 & 901 $\pm$ 901 \\
S40555/02 & PG1520+525 & 230.5 & 52.4 & 85.4 & 52.4 & 6183 & 4443 $\pm$ 3829 & 1216 $\pm$ 925 & 1022 $\pm$ 1022 & 3436 $\pm$ 2010 & 788 $\pm$ 788 & 1532 $\pm$ 1532 \\
S40557/03 & LSE44 & 208.2 & -48.1 & 313.4 & 13.5 & 3683 & 6048 $\pm$ 3058 & 4530 $\pm$ 1047 & 4423 $\pm$ 1313 & 4144 $\pm$ 1429 & 3613 $\pm$ 1826 & 3524 $\pm$ 1469 \\
S40557/01 & LSE44 & 208.2 & -48.1 & 313.4 & 13.5 & 18164 & 8089 $\pm$ 1331 & 5533 $\pm$ 375 & 4756 $\pm$ 828 & 6296 $\pm$ 1176 & 4003 $\pm$ 834 & 3940 $\pm$ 643 \\
S40557/02 & LSE44 & 208.2 & -48.1 & 313.4 & 13.5 & 6238 & 7571 $\pm$ 2538 & 4847 $\pm$ 646 & 4100 $\pm$ 1152 & 5588 $\pm$ 1575 & 3979 $\pm$ 1429 & 3462 $\pm$ 758 \\
S40557/04 & LSE44 & 208.2 & -48.1 & 313.4 & 13.5 & 10715 & 5084 $\pm$ 1571 & 5463 $\pm$ 591 & 4807 $\pm$ 835 & 5971 $\pm$ 1199 & 2787 $\pm$ 821 & 3085 $\pm$ 642 \\
S40558/01 & HD102567 & 177.0 & -62.2 & 295.6 & -0.2 & 7395 & 8829 $\pm$ 4827 & 6690 $\pm$ 1181 & 6361 $\pm$ 1700 & 8071 $\pm$ 1857 & 7878 $\pm$ 1721 & 4706 $\pm$ 1389 \\
S40563/01 & WD1634-573 & 249.6 & -57.5 & 329.9 & -7.0 & 15616 & 4598 $\pm$ 3584 & 2873 $\pm$ 707 & 3452 $\pm$ 1069 & 4809 $\pm$ 1187 & 3172 $\pm$ 2104 & 1317 $\pm$ 866 \\
S40563/02 & WD1634-573 & 249.6 & -57.5 & 329.9 & -7.0 & 2704 & 5337 $\pm$ 4845 & 4367 $\pm$ 2012 & 4355 $\pm$ 1746 & 4123 $\pm$ 2647 & 11832 $\pm$ 9902 & 4751 $\pm$ 4534 \\
S40563/03 & WD1634-573 & 249.6 & -57.5 & 329.9 & -7.0 & 3422 & 5992 $\pm$ 5437 & 3615 $\pm$ 1101 & 4456 $\pm$ 2038 & 5645 $\pm$ 2422 & 4033 $\pm$ 4033 & 2432 $\pm$ 1850 \\
S40563/04 & WD1634-573 & 249.6 & -57.5 & 329.9 & -7.0 & 12029 & 6560 $\pm$ 4195 & 3468 $\pm$ 1154 & 2749 $\pm$ 907 & 3530 $\pm$ 1573 & 1888 $\pm$ 1138 & 1279 $\pm$ 1279 \\
S40564/01 & BD+43D4035 & 341.7 & 44.3 & 100.6 & -13.1 & 5062 & 2584 $\pm$ 2584 & 2083 $\pm$ 1406 & 1277 $\pm$ 1277 & 2382 $\pm$ 2140 & 2873 $\pm$ 2643 & 1072 $\pm$ 940 \\
S40573/01 & HD35580 & 80.6 & -56.1 & 264.2 & -34.5 & 10555 & 3442 $\pm$ 3223 & 2571 $\pm$ 1094 & 1970 $\pm$ 1173 & 3236 $\pm$ 1075 & 2293 $\pm$ 2293 & 837 $\pm$ 837 \\
S40573/01 & HD35580 & 80.6 & -56.1 & 264.2 & -34.5 & 10555 & 4314 $\pm$ 3490 & 2529 $\pm$ 1029 & 2329 $\pm$ 1042 & 4088 $\pm$ 1463 & 2431 $\pm$ 1840 & 780 $\pm$ 780 \\
S40578/01 & HD074194 & 130.2 & -45.1 & 264.0 & -2.0 & 9611 & 11045 $\pm$ 2313 & 20502 $\pm$ 609 & 8553 $\pm$ 1394 & 8112 $\pm$ 2008 & 2395 $\pm$ 1157 & 1807 $\pm$ 675 \\
S40584/01 & WD1725+586 & 261.7 & 58.6 & 87.2 & 33.8 & 3838 & 4191 $\pm$ 4191 & 2008 $\pm$ 1654 & 2631 $\pm$ 1613 & 3569 $\pm$ 2454 & 2623 $\pm$ 1330 & 950 $\pm$ 950 \\
S40590/01 & HE2-138 & 237.9 & -66.3 & 319.7 & -9.5 & 10839 & 2587 $\pm$ 2587 & 2749 $\pm$ 933 & 2664 $\pm$ 1154 & 3295 $\pm$ 1304 & 2217 $\pm$ 1648 & 2506 $\pm$ 1185 \\
S40591/01 & HD104994 & 181.3 & -62.1 & 297.6 & 0.3 & 3216 & 22837 $\pm$ 7282 & 12420 $\pm$ 1306 & 11671 $\pm$ 2954 & 13675 $\pm$ 3231 & 15238 $\pm$ 2553 & 13071 $\pm$ 2346 \\
\enddata
\tablecomments{Columns 1 to 6 give the surface brightness of the diffuse radiation observed
in the respective rows of Table 1. The units are in \phunit\ and the uncertainties are 90\%
confidence limts.}
\end{deluxetable}

\end{document}